\documentstyle[twoside,11pt]{article}
\normalsize
\def\greaterthansquiggle{\raise.3ex\hbox{$>$\kern-.75em\lower1ex\hbox{$\sim$}}}
\def\lessthansquiggle{\raise.3ex\hbox{$<$\kern-.75em\lower1ex\hbox{$\sim$}}}
\newcommand{\bdi}{\begin{displaymath}}
\newcommand{\edi}{\end{displaymath}}
\newcommand{\bfi}{\begin{figure}}
\newcommand{\efi}{\end{figure}}

\newcommand{\beq}{\begin{equation}}
\newcommand{\eeq}{\end{equation}}

\newcommand{\gaM}{\gamma^{\mu}}

\newcommand{\beqa}{\begin{eqnarray}}
\newcommand{\eeqa}{\end{eqnarray}}
\newcommand{\no}{\nonumber}

\newcommand{\ra}{\rightarrow}

\def\au{{\setbox0=\hbox{\lower1.36775ex%
\hbox{''}\kern-.05em}\dp0=.36775ex\hskip0pt\box0}}
\def\ao{{}\kern-.10em\hbox{``}}

\newcommand{\dsla}{\partial\hspace{-6pt} /  }

\newcommand{\ddsla}{\partial\hspace{-4.6pt} /  }

\newcommand{\AAsla}{A\hspace{-5pt}  /  }

\begin{document}

\begin{titlepage}
\begin{flushright}
BUTP-95/26
\end{flushright}
\begin{center}

{\Large Vacuum functional and fermion condensate in the massive Schwinger
model}\\

\bigskip

Christoph Adam \\
Institut f\"ur theoretische Physik, Universit\"at Bern \\
Sidlerstra\ss e 5, CH-3012 Bern, Switzerland$^*)$ \\

\bigskip

\today \\

\bigskip

{\bf Abstract} \\

\bigskip

 We derive a systematic procedure of computing the vacuum functional
and fermion condensate of the massive Schwinger model via a perturbative
expansion in the fermion mass. We compute numerical results for the first
nontrivial order.

\vfill

\end{center}
$^*)${\footnotesize permanent address: Institut f\"ur theoretische Physik,
Universit\"at Wien \\
Boltzmanngasse 5, 1090 Wien, Austria \\
email address: adam@pap.univie.ac.at}
\end{titlepage}

\section*{Introduction}
The massless Schwinger model has been studied extensively because, although
being exactly soluble, it has quite a rich structure that resembles features
of more realistic theories. Among these are: a massive physical state is
formed via the chiral anomaly (\cite{Sc1}, \cite{LS1}, \cite{DR1}, \cite{GS1},
\cite{Le1}, \cite{ABH}, \cite{Diss}, \cite{Adam})
(this state is noninteracting in the massless Schwinger model). There are
"instanton-like" gauge field configurations present, and therefore the vacuum
structure is nontrivial: the true vacuum is a superposition of all instanton
sectors, and each possible superposition is labelled by a vacuum angle
$\theta$ (\cite{LS1}, \cite{tH1}, \cite{Jay}, \cite{SW1}, \cite{Sm1},
\cite{Diss},
\cite{Adam}, \cite{DSEQ}). However the physics does not depend on the value of
$\theta$ in the massless case.

The massive Schwinger model is no longer exactly soluble (\cite{CJS},
\cite{Co1}, \cite{Fry}, \cite{ABH}). But all of the
nontrivial features of the massless model persist to hold, at least for small
fermion mass. The massive state is now interacting, and its mass acquires
corrections due to the fermion mass.
Instantons and nontrivial vacuum are present, too, and, in addition, physical
quantities now depend on the value of the vacuum angle $\theta$.

Here we will show how some physical quantities may be computed via a
perturbative expansion in the fermion mass, and we will arrive at some
numerical results for the vacuum functional and fermion condensate.
For this we use the wellknown exact path integral solution of the massless
Schwinger model as a starting point.

\section*{Perturbative computation}
The vacuum functional for the massive Schwinger model may be written as a sum
of all instanton sectors ($k\ldots$ instanton number)
\beq
Z(m,\theta)=\sum_{k=-\infty}^{\infty}e^{ik\theta}Z_k (m)
\eeq
where
\bdi
Z_k (m)=N\int D\bar\Psi D\Psi DA^\mu_k e^{\int dx\Bigl[ \bar\Psi(i\ddsla
-e\AAsla_k -m)\Psi -\frac{1}{4}F_{\mu\nu}F^{\mu\nu}\Bigr] }
\edi
\bdi
=N\int D\bar\Psi D\Psi D\beta_k \sum_{n=0}^\infty \frac{m^n}{n!}\prod_{i=1}^n
\int dx_i \bar\Psi (x_i)\Psi (x_i)\cdot
\edi
\beq
\cdot \exp\{\int dx\Bigl[ \bar\Psi (i\dsla -\epsilon_{\mu\nu}\gaM\partial^\nu
\beta_k
)\Psi +\frac{1}{2}\beta_k \Box^2 \beta_k \Bigr] \}
\eeq
($A_\mu =\epsilon_{\mu\nu}\partial^\nu \beta$ corresponding to Lorentz gauge)
where we expanded the mass term. Therefore the vacuum functional is related to
the computation of fermionic VEV of the massless Schwinger model (and in
addition some space time integrations thereof). These fermionic VEV of the
massless Schwinger model may be computed from the generating functional
\bdi
Z_k [\bar\eta ,\eta]=N\int D\beta \prod_{i_0 =0}^{k-1} (\bar\eta
\Psi_{i_0}^\beta )(\bar\Psi_{i_0}^\beta \eta)\cdot
\edi
\beq
\cdot e^{i\int dxdy\bar\eta (x)G^\beta (x,y)\eta (y)}e^{\frac{1}{2}\int
dx\beta {\rm\bf D}\beta}
\eeq
where the fermions have already been integrated out. $\Psi_{i_0}^\beta$ are
the fermionic zero modes and $G^\beta$ is the exact fermionic Green's function
in the external gauge field $\beta$, and {\bf D} is the operator of the
effective gauge field action, with Green's function {\bf G} ($D_\mu (x) (D_0
(x))$ is the massive (massless) scalar propagator)
\bdi
{\rm\bf G}(x)=\pi (D_\mu (x)-D_0 (x))
\edi
\beq
D_\mu (x)=-\frac{1}{2\pi}K_0 (\mu |x|) \quad ,\quad D_0 (x)=\frac{1}{4\pi}\ln
x^2
\eeq
($K_0 \ldots$ McDonald function).
 From this all fermionic VEV may be computed (for details see e.g. \cite{Adam},
\cite{Diss}, \cite{DSEQ}).

It is useful to rewrite the fermionic scalar bilinears in terms of chiral
bilinears like
\beq
S(x):=\bar\Psi (x)\Psi (x)=\bar\Psi (x)P_+ \Psi (x)+\bar\Psi (x)P_- \Psi (x)
=:S_+ (x) +S_- (x),
\eeq
because for a VEV of products of $S_\pm $ only a definite instanton sector
contributes. E.g. the fermion condensate is
\beq
\langle S(x)\rangle =\langle S_+ (x)\rangle^{k=1}+\langle S_-
(x)\rangle^{k=-1}=\frac{1}{\pi}e^{2{\rm\bf G}(0)}=\frac{e^\gamma}{2\pi}\mu
=:\Sigma
\eeq
where $\gamma$ is the Euler constant and $\mu$ the Schwinger mass $\mu^2
=\frac{e^2}{\pi}$.

For higher VEV of $n_+ S_+$ and $n_- S_-$ one finds
\beq
\langle S_{H_1}(x_1)\cdots S_{H_n}(x_n)\rangle =\Bigl( \frac{\Sigma}{2}\Bigr)^n
\exp
\Bigl[ \sum_{i<j}(-)^{\sigma_i \sigma_j}4\pi D_\mu (x_i -x_j)\Bigr]
\eeq
where $\sigma_i =\pm 1$ for $H_i =\pm$. The sole contribution to this VEV
stems from the instanton sector $k=\sum_i \sigma_i =n_+ -n_- $ (see e.g.
\cite{Adam}, \cite{DSEQ}, \cite{Zah}).

Now we just have to insert this result into the perturbation expansion (2).
For a first, rough approximation we may use the fact that the massive scalar
propagator $D_\mu (x)$ vanishes exponentially for large argument. Therefore,
when integrating over space time and expanding the exponential, all
contributions from $D_\mu^l $ will be ignored for the moment, supposing that
the space time volume $V$ is sufficiently large. In this case the integrations
in (2) just produce factors of $V$. Further, when inserting (7) in (2) we
have to sum over all possible distributions of $n_+ =n-n_-$ pluses and $n_-$
minuses on $n$ scalar densities $S$. This results in a factor $n \choose n_-$.
Therefore we find for the $n$-th order term
\beq
\langle\prod_{i=1}^n \int dx_i S(x_i )\rangle \sim \Bigl(
\frac{\Sigma}{2}\Bigr) ^n V^n
\sum_{n_- =0}^n {n \choose n_-}e^{i(n-2n_-)\theta}=(\Sigma V\cos \theta )^n
\eeq
and for the normalized vacuum functional
\beq
\frac{Z(m,\theta)}{Z(0,0)}\sim \exp (m\Sigma V\cos\theta)
\eeq
(we ignored terms like $m^n V^{n-1}$ in this approximation compared to
$m^{n-1}V^{n-1}$, therefore (9) is the first order result in $m$). This
result is wellknown, and its consequences for the vacuum structure and
spectrum of the Dirac operator are discussed in great detail in \cite{LSm}.

To obtain higher order results, we rewrite the exponential in (7) like e.g.
\bdi
\exp \Bigl( \sum (-)^{\sigma_i \sigma_j}4\pi D_\mu (x_i -x_j)\Bigr) =
\exp \Bigl( +4\pi D_\mu
(x_1 -x_2)\Bigr) \cdot \exp\Bigl( -4\pi D_\mu (x_1 -x_3)\Bigr) \cdot\ldots
\edi
\beq
=\Bigl( 1+E(x_1 -x_2)\Bigr) \cdot \Bigl( 1+F(x_1 -x_3)\Bigr) \cdot \ldots
\eeq
where
\bdi
E(x)=e^{4\pi D_\mu (x)}-1\quad ,\quad F(x)=e^{-4\pi D_\mu (x)} -1
\edi
\beq
E\equiv \int d^2 xE(x)\quad ,\quad F\equiv \int d^2 xF(x).
\eeq
Both functions $E(x),F(x)$ vanish exponientially for large argument.

Using this notation and inserting (7) into the perturbation expansion (2) we
obtain, order by order: \\ \\
n=1:
\beq
\frac{m}{1!}\frac{\Sigma}{2}\int dx(e^{i\theta}+e^{-i\theta})=
\frac{m}{1!}\frac{\Sigma}{2}V(e^{i\theta}+e^{-i\theta})
\eeq
n=2:
\bdi
\frac{m^2}{2!}\Bigl( \frac{\Sigma}{2}\Bigr)^2 \int dx_1 dx_2
\Bigl[ e^{2i\theta}e^{4\pi D_\mu
(x_1 -x_2)}+2e^{-4\pi D_\mu (x_1 -x_2)}+e^{-2i\theta}e^{4\pi D_\mu (x_1
-x_2)}\Bigr]
\edi
\beq
=\frac{m^2}{2!}\Bigl( \frac{\Sigma}{2}\Bigr)^2 \Bigl[ V^2 (e^{2i\theta}+2
+e^{-2i\theta})+
V(Ee^{2i\theta}+2F+Ee^{-2i\theta})\Bigr]
\eeq
n=3:
\bdi
\ldots =\frac{m^3}{3!}\Bigl( \frac{\Sigma}{2}\Bigr)^3 \Bigl[ V^3
(e^{3i\theta}+3e^{i\theta}+3e^{-i\theta}+e^{-3i\theta})+
\edi
\bdi
V^2 \Bigl( 3Ee^{3i\theta}+3(E+2F)e^{i\theta}+
3(E+2F)e^{-i\theta}+3Ee^{-3i\theta}\Bigr) +
\edi
\bdi
V\Bigl( (3E^2 +E\times E\times E)e^{3i\theta}+3(2EF +F^2 +E\times F\times
F)e^{i\theta}+
\edi
\beq
3(2EF+F^2 +E\times F\times F)e^{-i\theta}+(3E^2 +E\times E\times E)
e^{-3i\theta}\Bigr) \Bigr]
\eeq
n=4:
\bdi
\ldots =\frac{m^4}{4!}\Bigl( \frac{\Sigma}{2}\Bigr)^4 \Bigl[ V^4
(e^{4i\theta}+4e^{2i\theta}
+6+4e^{-2i\theta}+e^{-4i\theta})+
\edi
\bdi
V^3 \Bigl( 6Ee^{4i\theta}+12(E+F)e^{2i\theta}+12(E+2F)+12(E+F)e^{-2i\theta}
+6Ee^{-4i\theta}\Bigr) +
\edi
\bdi
V^2 \Bigl( (15E^2 +4E\times E\times E)e^{4i\theta}+4(3E^2 +9EF+3F^2 +E\times
E\times
E+3E\times F\times F)e^{2i\theta}+
\edi
\bdi
+6(E^2 +8EF+F^2 +4E\times F\times F) + 4(3E^2 +9EF+3F^2+E\times E\times E
+3E\times F\times F)e^{-2i\theta}+
\edi
\beq
+(15E^2 +4E\times E\times E)e^{-4i\theta}\Bigr) +\ldots \Bigr]
\eeq
\bdi
\ldots
\edi
where e.g.
\beq
E\times E\times E\equiv \int dy_1 dy_2 E(y_1)E(y_1 +y_2)E(y_2)
\eeq
and we displayed the result up to the accuracy we need.
Observe that the result is not obtained by just expanding polynomials like
$(1+E(x_i)^l$, because e.g. a third power in $E(x_i)$ may contribute to
$V^{n-3}E^3$ or to $V^{n-2}E\times E\times E$.

In a next step we rearrange the terms (12) -- (15) in rising powers of $V$:
\bdi
\frac{V}{1!}\Bigl[ m\frac{\Sigma}{2}(e^{i\theta}+e^{-i\theta})+\frac{m^2}{2}
\Bigl( \frac{\Sigma}{2}\Bigr)^2 (Ee^{2i\theta}+2F+Ee^{-2i\theta})+
\edi
\bdi
\frac{m^3}{6}\Bigl( \frac{\Sigma}{2}\Bigr)^3 \Bigl( (3E^2 +E\times E\times
E)(e^{3i\theta}+
e^{-3i\theta})+3(2EF+F^2 +E\times F\times F)(e^{i\theta}+e^{-i\theta})\Bigr)
+\ldots\Bigr] +
\edi
\bdi
\frac{V^2}{2!}\Bigl[ m^2 \Bigl( \frac{\Sigma}{2}\Bigr)^2
(e^{2i\theta}+2+e^{-2i\theta})+
m^3 \Bigl( \frac{\Sigma}{2}\Bigr)^3
\Bigl( E(e^{3i\theta}+e^{-3i\theta})+(E+2F)(e^{i\theta}
+e^{-i\theta})\Bigr) +
\edi
\bdi
\frac{m^4}{12}\Bigl( \frac{\Sigma}{2}\Bigr)^4 \Bigl(
(15E^2 +4E\times E\times E)(e^{4i\theta}+e^{-4i\theta})+
\edi
\bdi
4(3E^2 +9EF+3F^2 +E\times E\times E +3E\times F\times F)(e^{2i\theta}+
e^{-2i\theta})+
\edi
\bdi
6(E^2 +8EF+F^2 +4E\times F\times F)\Bigr) +\ldots\Bigr] +
\edi
\bdi
\frac{V^3}{3!}\Bigl[ m^3 \Bigl( \frac{\Sigma}{2}\Bigr)^3 (e^{3i\theta}+
3e^{i\theta}+3e^{-i\theta}+e^{-3i\theta})+
\edi
\bdi
\frac{m^4}{2}\Bigl( \frac{\Sigma}{2}\Bigr)^4 \Bigl( 3E(e^{4i\theta}+
e^{-4i\theta})+
6(E+F)(e^{2i\theta}+e^{-2i\theta})+(E+2F)\Bigr) +\ldots \Bigr] +
\edi
\bdi
\frac{V^4}{4!}\Bigl[ m^4 \Bigl( \frac{\Sigma}{2}\Bigr)^4
(e^{4i\theta}+4e^{2i\theta}+6
+4e^{-2i\theta}+e^{-4i\theta})+\ldots \Bigr] +\ldots
\edi
\beq
=:\frac{V}{1!}\alpha +\frac{V^2}{2!}\alpha^2 +\frac{V^3}{3!}\alpha^3 +\ldots
\eeq
We find that the coefficient of $\frac{V^n}{n!}$ is the $n$-th power of some
specific function $\alpha$ where $\alpha$ does not depend on $V$ any more.
This feature could have been expected, because the fermion condensate should
not depend on the volume $V$. With this function $\alpha$,
\bdi
\alpha (m,\theta )=m\frac{\Sigma}{2}2\cos \theta
+\frac{m^2}{2!}\Bigl( \frac{\Sigma}{2}\Bigr)^2 (2E\cos 2 \theta +2F)+
\edi
\beq
\frac{m^3}{3!}\Bigl( \frac{\Sigma}{2}\Bigr) ^3 \Bigl( (3E^2 +E\times E\times
E)2\cos 3\theta
+3(2EF+F^2 +E\times E\times E)2\cos \theta \Bigr) +\ldots
\eeq
the normalized vacuum functional may be written like
\beq
\frac{Z(m,\theta )}{Z(0,0)}=e^{V\alpha (m,\theta)}.
\eeq
 From this it is very easy to compute the fermion condensate
\beq
\langle\bar\Psi\Psi\rangle (m,\theta)\equiv\frac{1}{V}\frac{\partial}{\partial
m}
\ln Z(m,\theta )=\frac{\partial}{\partial m}\alpha (m,\theta)
\eeq
\bdi
\langle\bar\Psi\Psi\rangle (m,\theta )=\Sigma\cos\theta +\frac{m}{2}\Sigma^2
(E\cos 2\theta +F)+
\edi
\beq
\frac{m^2}{8}\Sigma^3 \Bigl( (3E^2 +E\times E\times E)\cos 3\theta +
3(2EF+F^2 +E\times F\times F)\cos \theta \Bigr) +\ldots
\eeq
For a numerical evaluation of order $m$ we have to compute the coefficients
$E$ and $F$. First, both $E$ and $F$ are proportional to $\frac{1}{\mu^2}$.
For $\mu =1$, $E$ is
\beqa
E=\int d^2 xE(x) &=& \int d^2 x(e^{-2K_0 (|x|)}-1) \no \\
&=& 2\pi \int_0^\infty drr(e^{-2K_0 (r)}-1) = -8.9139
\eeqa
$E(x)$ is well behaving ($E(0)=-1$), so the numerical integration is straight
forward. $F(x)$ is singular at $x=0$, $F(x)\sim \frac{1}{x^2}$ for $x\to 0$,
but this singularity can easily be understood and removed in a unique way.
Indeed, this singularity is just the free fermion singularity, as can be seen
by rewriting $F(x)$ like
\beq
\Sigma^2 (F(x)+1)=G_0^2 (x)e^{4\pi ({\rm\bf G}(0)-{\rm\bf G}(x))}
\stackrel{|x| \ra 0}{\longrightarrow} G_0^2 (x)=\frac{1}{4\pi^2 x^2}.
\eeq
This singularity may be isolated by a partial integration:
\beqa
F &=& \int d^2 x(e^{2K_0 (|x|)}-1)=2\pi \int_0^\infty \frac{dr}{r}
(e^{2K_0 (r)+2\ln r}-r^2) \no \\
&=& 2\pi\Bigl[ \ln r(e^{2K_0 (r)+2\ln r}-r^2)\Bigr]_{\epsilon\to 0}^\infty \no
\\
&+& 2\pi\int_0^\infty dr2\ln r((K_1 (r)-\frac{1}{r})e^{2K_0 (r)+2\ln r} + r)
\eeqa
($K_0^{'} =-K_1$). Observe that the first term precisely leads to the free
field
singularity at the lower boundary (and vanishes at the upper boundary). So the
second term is the unique and finite result we are looking for. The numerical
integration gives
\beq
F=9.7384
\eeq
With these results we find for the fermion condensate
\beq
\langle\bar\Psi\Psi\rangle (m,\theta )=\Sigma\cos \theta
+\frac{m}{2}\frac{\Sigma^2}{\mu^2} (-8.9139 \cos 2\theta + 9.7384) +o(m^2)
\eeq
or, using (6),
\beqa
\frac{1}{\mu}\langle\bar\Psi\Psi\rangle (m,\theta ) &=& \frac{e^\gamma}{2\pi}
\cos\theta +\frac{m}{\mu}\frac{e^{2\gamma}}{8\pi^2}(E\cos 2\theta +F)
+o(\frac{m^2}{\mu^2}) \no \\
&=& 0.2835 \cos\theta +\frac{m}{\mu}(0.7825 - 0.7163 \cos 2\theta )+
o(\frac{m^2}{\mu^2}).
\eeqa
This is our final result. Observe that the correction is minimal for $\theta
=0$.

\section*{Summary}
We have derived a systematic procedure of computing the vacuum functional and
fermion condensate of the massive Schwinger model in a mass perturbation
theory, order by order. For the first nontrivial order we even displayed
numerical results. For higher orders the numerical calculations are more
involved (because of the occurrence of linked functions in the integrations,
like $E\times E\times E$), but in principle they could be performed, e.g. in
order to compare the result to the results of other approaches.

The numerical computations in this article were done with Mathematica 2.2.

\section*{Acknowledgements}
The author thanks H. Leutwyler for very helpful discussions and the members of
the Institute of Theoretical Physics at Bern University, where this work was
done, for their hospitality.

This work was supported by a research stipendium of the University of Vienna.

\end{document}